\documentclass[amsthm]{elsart}

\usepackage{epsfig}

\usepackage{amssymb}

\begin{document}

\begin{frontmatter}

\title{On sensitivity calculations for neutrino oscillation experiments}

\author[label1]{Jan Conrad}

\address[label1]{Royal Institute of Technology (KTH) and High Energy Astrophysics and Cosmology Centre (HEAC),
AlbaNova University Centre, 10691 Stockholm, Sweden}

\begin{abstract}
Calculations of sensitivities of future experiments are a necessary ingredient in experimental high energy physics. Especially in the context of measurements of the neutrino oscillation parameters extensive studies are performed to arrive at the optimal configuration. In this note we clarify the definition of sensitivity as often applied in these studies. In addition we examine two of the most common methods to calculate sensitivity from a statistical perspective using a toy model. The importance of inclusion of uncertainties in nuisance parameters for the interpretation of sensitivity calculations is pointed out.
\end{abstract}


\end{frontmatter}

\section{Introduction}
In the process of developing experiments measuring new phenomena in physics the estimation of the sensitivity of certain configurations of experiments is of utmost importance.\\

\noindent
A particularly active field is the estimate of sensitivities for future neutrino oscillation experiments, see for example \cite{ISS:2007}. The goal of future experiments is often the measurement of a mixing angle (denoted $\Theta_{13}$), the probability for a neutrino oscillation taking place being proportional to $\sin^2{2\Theta_{13}}$.\\

\noindent
In sensitivity studies for neutrino oscillation experiments, ``sensitivity'' is often not defined in the same way. Furthermore uncertainties in nuisance parameters (often sloppily called ``systematic uncertainties''\footnote{we refer the reader to \cite{Sinervo:2003wm} for a discussion on how to classify uncertainties in nuisance parameters}) are often ignored (see for example \cite{Komatsu:2002sz,Indumathi:2006gr}) or included in calculations in an incomplete manner (see discussion in \cite{Conrad:2006}). Little attention seems to be given to the issue of how sensitivity is defined and uncertainties are treated, despite the fact that decisions on experimental set-ups might be based on small differences in sensitivity studies.\\

\noindent
In this note we try to firstly clarify the definition of ``sensitivity'' and discuss - using a Toy model - potential problems which arise if instrumental uncertainties need to be considered.\\ 

\noindent
The issue of sensitivity calculation has after being discussed on a recent conference on future neutrino experiments \cite{Conrad:2006} already inspired a more careful assessment of sensitivity calculation \cite{Schwetz:2006md} which indicates that a more formal discussion is worthwhile.

\section{Definitions of sensitivity}
Probably the most common definition of sensitivity adopted in the study of future experiments aimed at the discovery of signals of as yet undetected physics phenomena is:\\ 

\noindent
{\it   The experiment is said to be sensitive to a given  value of   the parameter $\Theta_{13} = \Theta_{13}^{sens}$  at significance level $\alpha$ if the mean p-value obtained given $\Theta_{13}^{sens}$ is smaller than $\alpha$.}\\

\noindent Here we choose (in the spirit of neutrino oscillation experiments) the parameter describing the new physics phenomenon to be denoted by $\Theta_{13}$. The p-value is (per definition) calculated under the condition that the null hypothesis holds:
\begin{equation}
p = P(T \geq t_{obs} | H_{0}:\Theta_{13} = 0 )
\end{equation}
where $H_{0}$ denotes the null hypothesis and $T$ denotes the test statistics with its observed value $t_{obs}$, which is distributed as the distribution function $P$. We will give common definitions of $T$ in the next section.\\

\noindent
A variation, which is the most commonly used in the context of neutrino experiments is using confidence intervals for the definition of sensitivity:\\

\noindent
{\it The experiment is said to be sensitive to a given value of the parameter $\Theta_{13} = \Theta_{13}^{sens}$  at significance level $\alpha$ if the mean $1-\alpha$ confidence interval obtained, given $\Theta_{13}^{sens}$ , does not contain $\Theta_{13} = 0$}\\

\noindent
Both definitions can be equivalent, but are not in general, depending on the choice of test statistics and the method of confidence interval calculation. For example, choosing to calculate upper limits would never yield detection.\\

\noindent
To our knowledge, all sensitivity curves presented for neutrino oscillation experiments follow above definitions and therefore correspond to the mean observation. This has two important implications: firstly, it should be emphasized that if the distribution of parameter estimates is Gaussian, it is a well known fact (but often ignored) that even if $\Theta_{13}^{true} = \Theta_{13}^{sens}$ (i.e. the true value is at the estimated sensitivity), the probability for actually claiming discovery is only 50 \%. Secondly, if the parameter estimates are not Gaussian (see for example \cite{Schwetz:2006md}), then the presentation of the mean result yields very little information on the actual probability for discovery. Thus, the choice to present mean experimental results  is not  particularly informative. A more general definition of sensitivity therefore has to specify two probabilities:\\

\noindent
{\it The experiment is sensitive to a given value of $\Theta_{13} = \Theta_{13}^{sens}$  if the probability of obtaining an observation n which rejects $\Theta_{13} =0$ with at least significance $\alpha$ is at least $\beta$.}\\

\noindent
where we have chosen to reformulate the definition as to be applicable to both definitions given above. Another example where an attempt is made to define sensitivity in a manner involving the detection probability can be found in \cite{Punzi:2003bu}

\section{Claiming discovery and calculating confidence intervals}
According to the Neyman-Pearson lemma, the uniformly most powerful test statistics that can be chosen is the likelihood ratio:
\begin{equation}
T = \frac{\mathcal{L}(n|H_{0})}{\mathcal{L}(n|H_{1})}
\end{equation}
where T denotes the test statistics and $\mathcal{L}(n|H)$ denotes the likelihood under the observation $n$ for the null hypothesis ($H_0$) and the alternative hypothesis ($H_1$), respectively. One useful property of the likelihood ratio is the fact that asymptotically:
\begin{equation}
-2 \ln{T} \sim \chi^2 
\end{equation}
i.e. the distribution of T under the null hypothesis is known and the significance of the observation can be calculated from the $\chi^2$ distribution.
A particular common method in studies of neutrino oscillation experiments is to perform a $\chi^2$ fit and calculate a confidence interval from the function $\chi^2(\Theta_{13})$. For example, the interval $[\Theta_{13}^{low},\Theta_{13}^{up}]$ can be found by finding the points for which:
\begin{equation}
\chi^2(\Theta_{13}) - \chi^2_{min} = 2.706 
\end{equation}
where $\chi^2_{min}$ denotes the $\chi^2$ at the best fit value of $\Theta_{13}$. This confidence interval is then often used to claim discovery by requiring $\Theta_{13}^{low} > 0$. \\

\noindent
There are two quantities which are of crucial importance in the context of calculation of confidence intervals and in the testing of hypothesis (claiming of discovery). Methods to calculate confidence intervals should have {\it coverage}, defined as:\\

\noindent
{\it
An algorithm is said to have the correct {\em coverage\/} if given a confidence level $1-\alpha$ and a large number of repeated identical experiments, the resulting confidence intervals include the true value of the parameters to be estimated in a fraction $1-\alpha$ of all experiments.}\\

\noindent
If confidence intervals are used to claim discovery (meaning for testing hypotheses), then $\alpha$ is the probability for making a type I error, i.e. the probability for rejecting the null hypothesis though it is true (often called {\it significance}).\\

\noindent
The other quantity we will be interested in is the {\it power}. Power is the probability that the null hypothesis is rejected given that the alternative hypothesis is true. This quantity is exactly the probability we denoted $\beta$ in the previous section. The probability 1 - $\beta$ is the probability to make a type II error (accepting the null hypothesis though it is false).

\section{Including uncertainties in nuisance parameters}

{\it Nuisance parameters } are parameters which enter the data model, but which are not of prime interest. The probably most common example is the expected background in a Poisson process. Sensitivities (as confidence intervals) are usually only calculated for the parameter of primary interest and it is not desired to calculate them depending on parameters which are of no physical interest and specific to the experiment. Thus, ways have to be found to marginalize the nuisance parameter. There are two particularly common approaches: \\
\noindent 
In the first method, the probability density function (PDF) without uncertainty in nuisance parameters is replaced by one where there is an integration over all possible true values of the nuisance parameter {\it(integration method)}:
\begin{equation}
P(n|s,b_{true}) \longrightarrow \int_0^{\infty}{P(n|s,b_{true}) P(b_{true}|b_{est}) d\,b_{true}}
\label{eq:integration}
\end{equation}
Here $b_{true}$ is the true value of the nuisance parameter and $b_{est}$ is its estimate. Since the integrated PDF is describing the probability of the true value given its estimate (and not vice versa) this method is Bayesian. Some prior probability distribution of the true value of the nuisance parameter has to be assumed.\\
\noindent
In the other common method, the PDF is replaced by one where for each $s$ the PDF is maximized with respect to the nuisance parameters {\it(profiling method)}
\begin{equation}
P(n|s,b_{true}) \longrightarrow \max_{b_{true}}{\mathcal{L}(n|s,b_{true})}
\label{eq:profile}
\end{equation}
with notation as above. This method is completely frequentist, since it never treats $b_{true}$ as a random variable. Therefore the argument of the maximization is a likelihood function and not a PDF. Both these methods are frequently applied in high energy physics in confidence interval calculations \cite{Conrad:2002kn},\cite{Rolke:2004mj} and references therein and to them. \\

\noindent
In assessment of sensitivities of neutrino oscillation experiments, uncertainties are often included by performing a least square fit using a  modified $\chi^2$ and use the resulting confidence interval (see previous section) to claim discovery.  Two modifications are particularly common. One method is to add the uncertainty in the background estimate in quadrature (for simplicity we will restrict ourselves to background estimate uncertainties):
\begin{equation}
\label{eq:chi11}
\chi^2_{add} = \frac{(n_{obs}- b_{est})^2}{b_{est} + \sigma_b^2}
\end{equation}
where $n_{obs}$ denotes the experimental result, $b_{est}$ the background estimate and $\sigma_b^2$ the uncertainty on that estimate.
Under assumption of a Gaussian process and applying Bayesian reasoning, $\chi^2_{add}$ can be viewed as equivalent to using the method illustrated in equation \ref{eq:integration}.\\

\noindent
The other (probably more common) method of inclusion is based on adding a normalization parameter to the $\chi^2$ and minimize the $\chi^2$ with respect to it, see for example \cite{Burguet-Castell:2005pa,Huber:2002mx,Barger:2006vy}
\begin{equation}
\label{eq:chi21}
\chi^2_{prof} = \min_A{ \left( \frac{(n_{obs}- Ab_{est})^2}{Ab_{est}} + \frac{(A-1)^2}{\sigma_{a}^2} \right) }
\end{equation}
where in addition to the parameters described above, we introduce the normalization parameter $A$. This modification is equivalent to the method represented by equation \ref{eq:profile} under assumption of Gaussian processes.\\

\noindent
A priori it can not be assumed that the modified quantities $\chi^2_{prof}$ and $\chi^2_{add}$ still follow a $\chi^2$ distribution. However, in general, this is the assumption employed in sensitivity calculations\footnote{If uncertainties are ignored, the $\chi^2$ is obviously not modified, in our simple example: $\chi^2 = \frac{(n-b_{est})^2}{b_{est}}$. However, its distribution under the null hypothesis is not $\chi^2$ since $b_{est}$ is not constant at the true value, but a random variable.}\footnote{The only exception known to us is \cite{Schwetz:2006md}}. In the following section we will apply above definitions to a toy model and check the validity of the assumption using Monte Carlo simulations. 

\section{Testing the $\chi^2$ method with a Toy model}

For simplicity, we will consider a one bin measurement, where we measure a number of events from a Poisson process with background contribution and we obtain an estimate of the background from a separate measurement, which is assumed to be Gaussian. In equations:
\begin{equation}
n \sim Po(s+b);\,\,\,\,\, b_{est} \sim G(b,\sigma_b)
\end{equation}
where $Po(s+b)$ denotes a Poisson process with experimental outcome $n$ (number of events), signal parameter $s$ and background parameter $b$ and $G(b,\sigma_b)$ denotes a Gaussian process with experimental outcome $b_{est}$ and width $\sigma_b$.
Since in the common neutrino oscillation experiment $s \propto \sin^2{2\Theta_{13}}$ this Toy model captures the main feature of many experiments (though being a simplification, obviously).\\

\noindent
Using Monte Carlo simulations of replica of the actual experiment, we  can calculate the true distribution of the test statistics defined in equation \ref{eq:chi11} and \ref{eq:chi21} under the condition that the null hypothesis is true ($s=0)$, thus the coverage. 
We can also assume $s > 0$ and calculate the probability that the null distribution will be rejected given the alternative hypothesis is true, i.e. the power.\\

\begin{center}
\begin{figure}
\epsfig{file=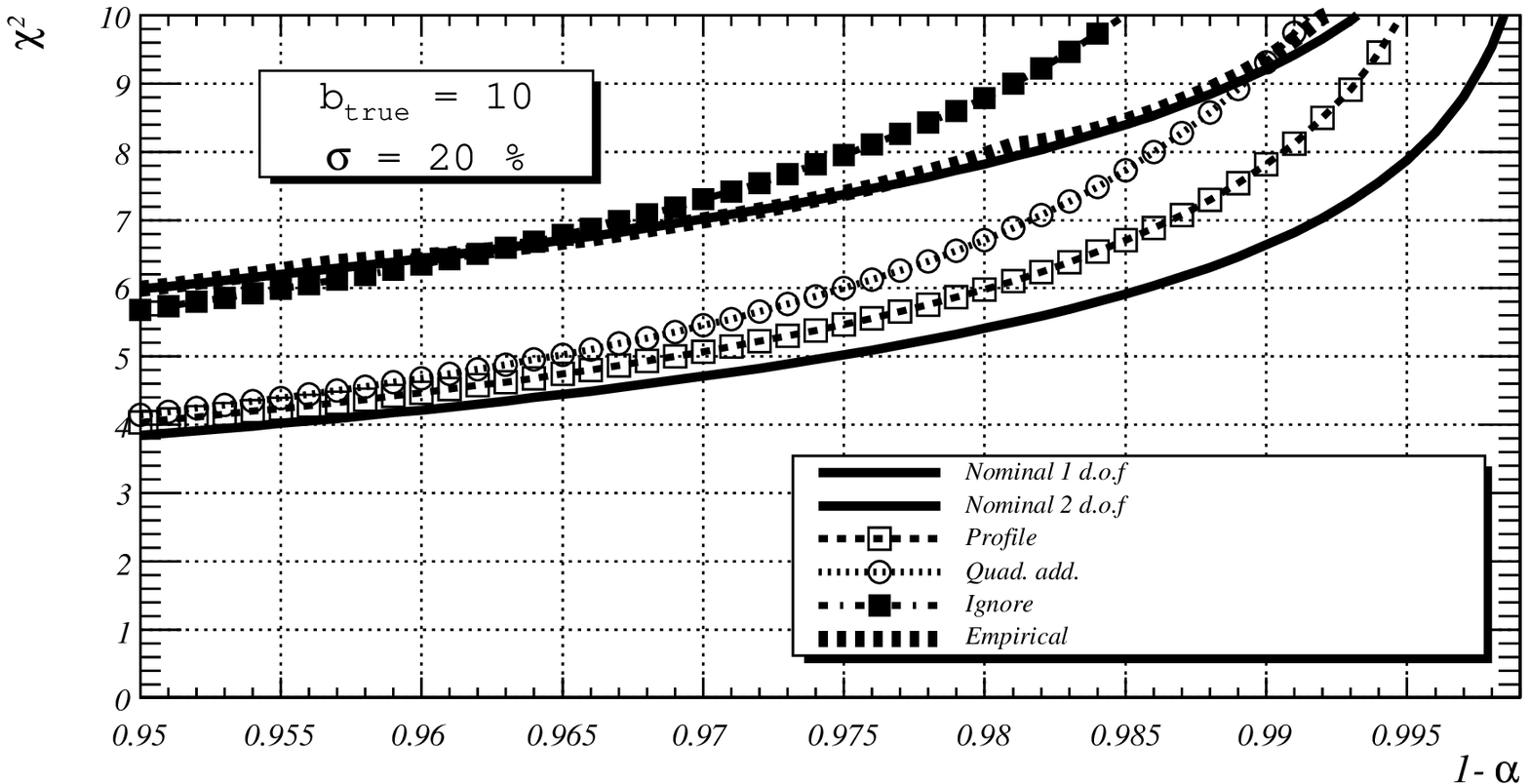,height=8cm,width=8cm} 
\epsfig{file=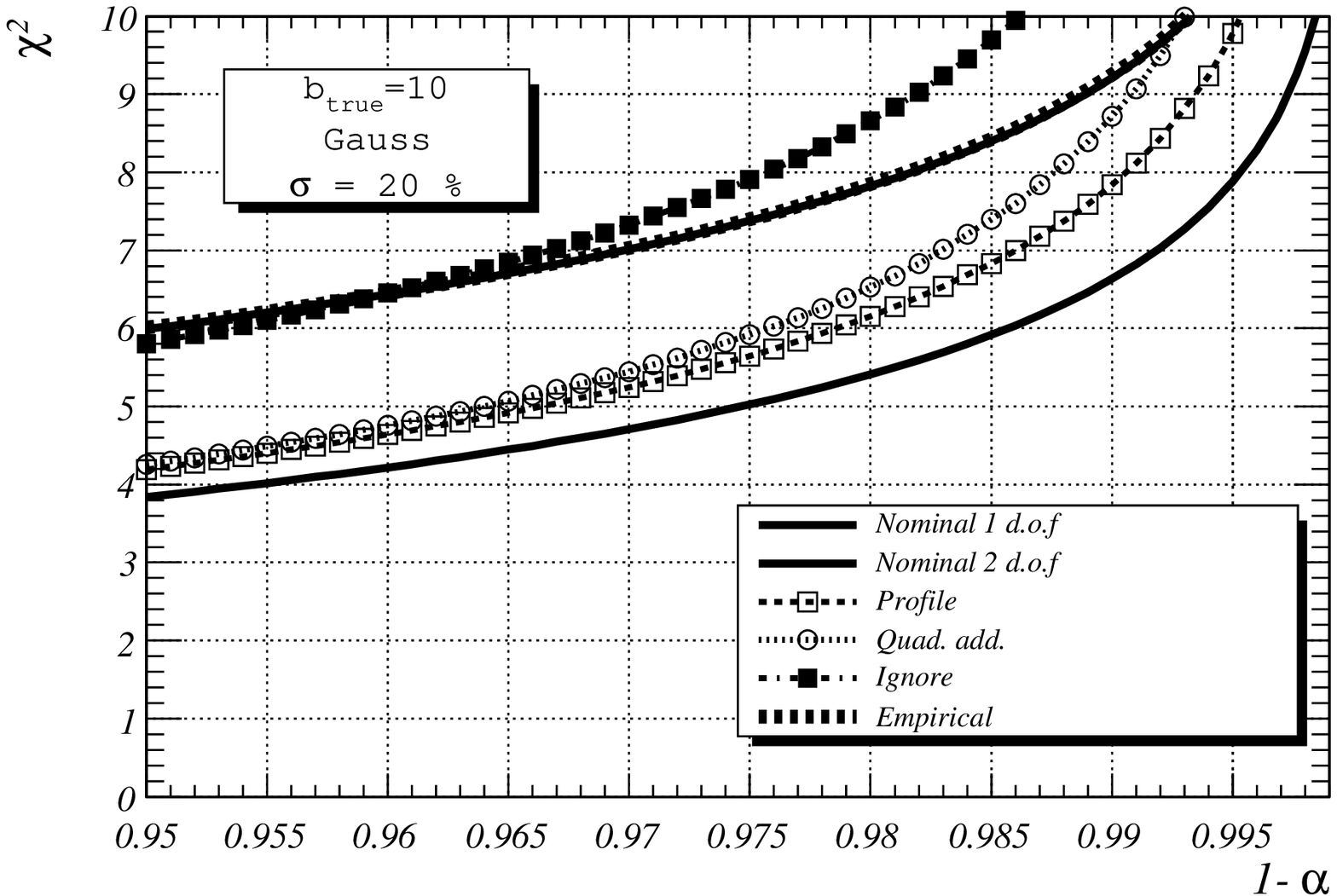,height=8cm,width=8cm}
\caption{Quantiles of the distribution of $\chi^2$ (Nominal 1 d.o.f.), $\chi^2_{add}$ (Quad. add.), $\chi^2_{prof}$ (Profile) and the $\chi^2$ where uncertainties in the background are ignored (Ignore). The curve labeled ``empirical'' shows the $\chi^2$ where both the number of events and the background are considered as measurements.}%
The left panel shows the results for the Poisson measurement process. The right panel assumes a Gaussian process.
The uncertainties in the background estimate are assumed to be 20 \%.
\label{fig:cov}
\end{figure}
\end{center}

\begin{center}
\begin{figure}
\epsfig{file=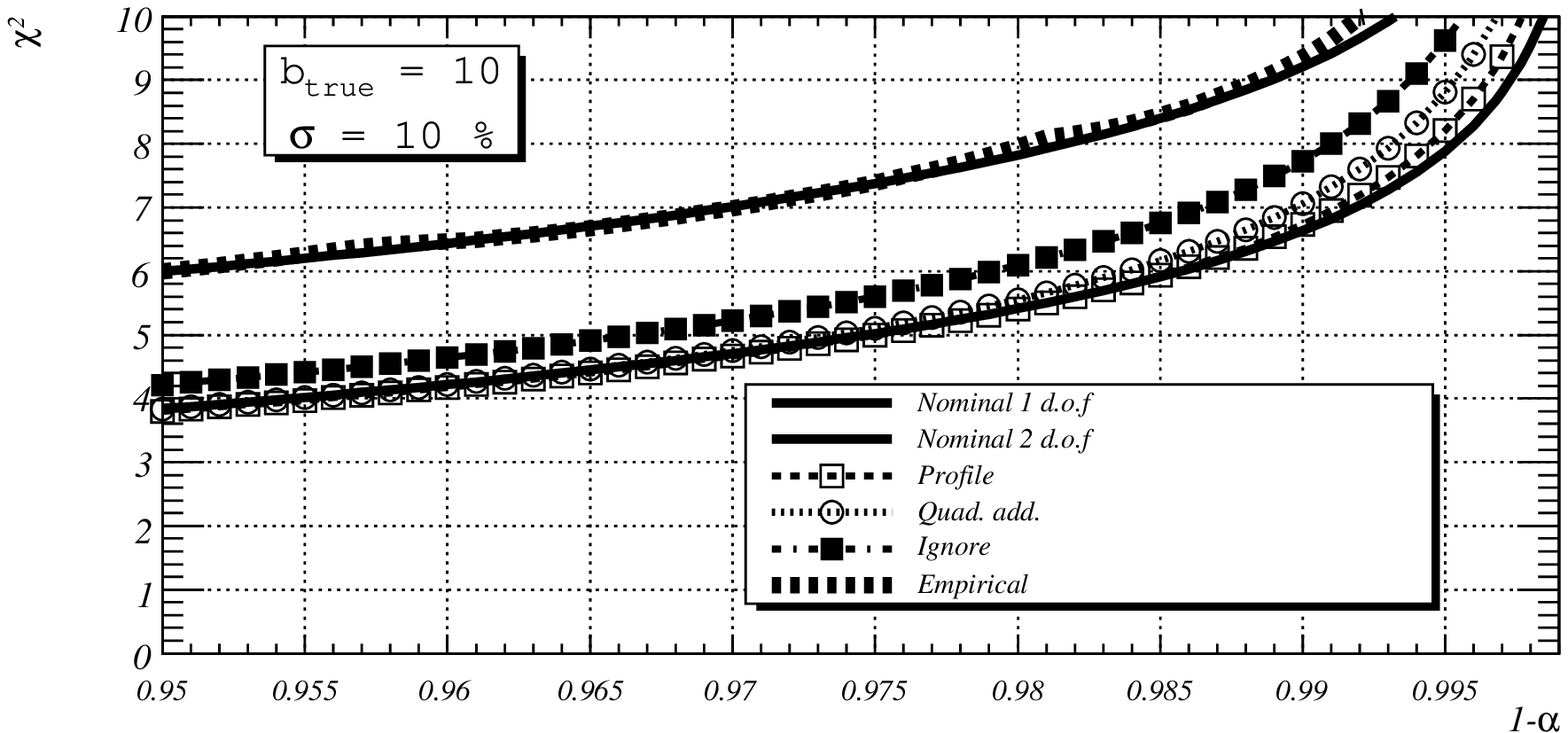,height=8cm,width=8cm} 
\epsfig{file=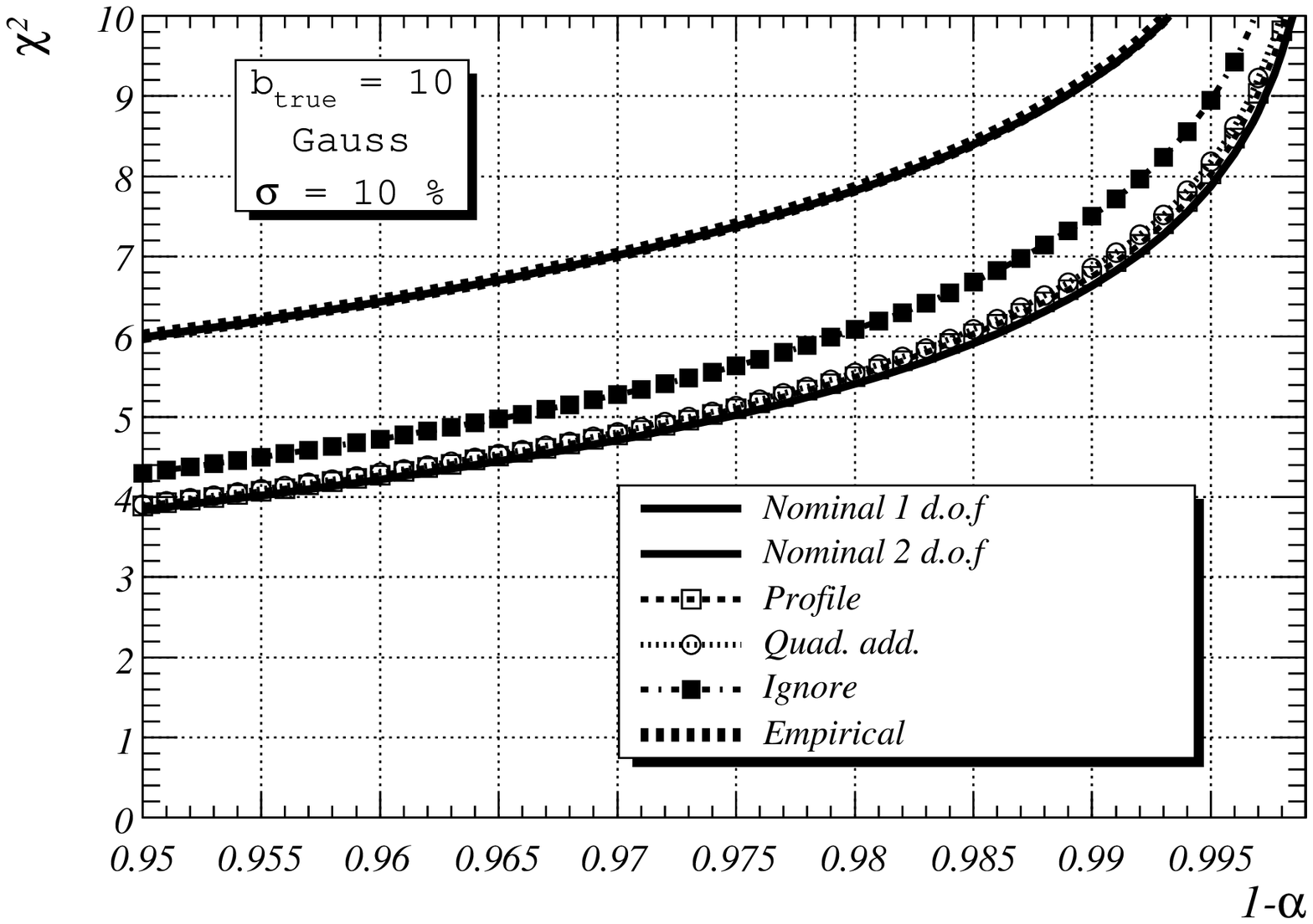,height=8cm,width=8cm}
\caption{Quantiles of the distribution of $\chi^2$ (Nominal 1 d.o.f.), $\chi^2_{add}$ (Quad. add.), $\chi^2_{prof}$ (Profile) and the $\chi^2$ where uncertainties in the background are ignored (Ignore). The curve labeled ``empirical'' shows the $\chi^2$ where both the number of events and the background are considered as measurements.
The left panel shows the results for the Poisson measurement process. The right panel assumes a Gaussian process. The uncertainties in the background estimate are assumed to be 10 \%.
}%

\label{fig:cov2}
\end{figure}
\end{center}

\begin{center}
\begin{figure}
\epsfig{file=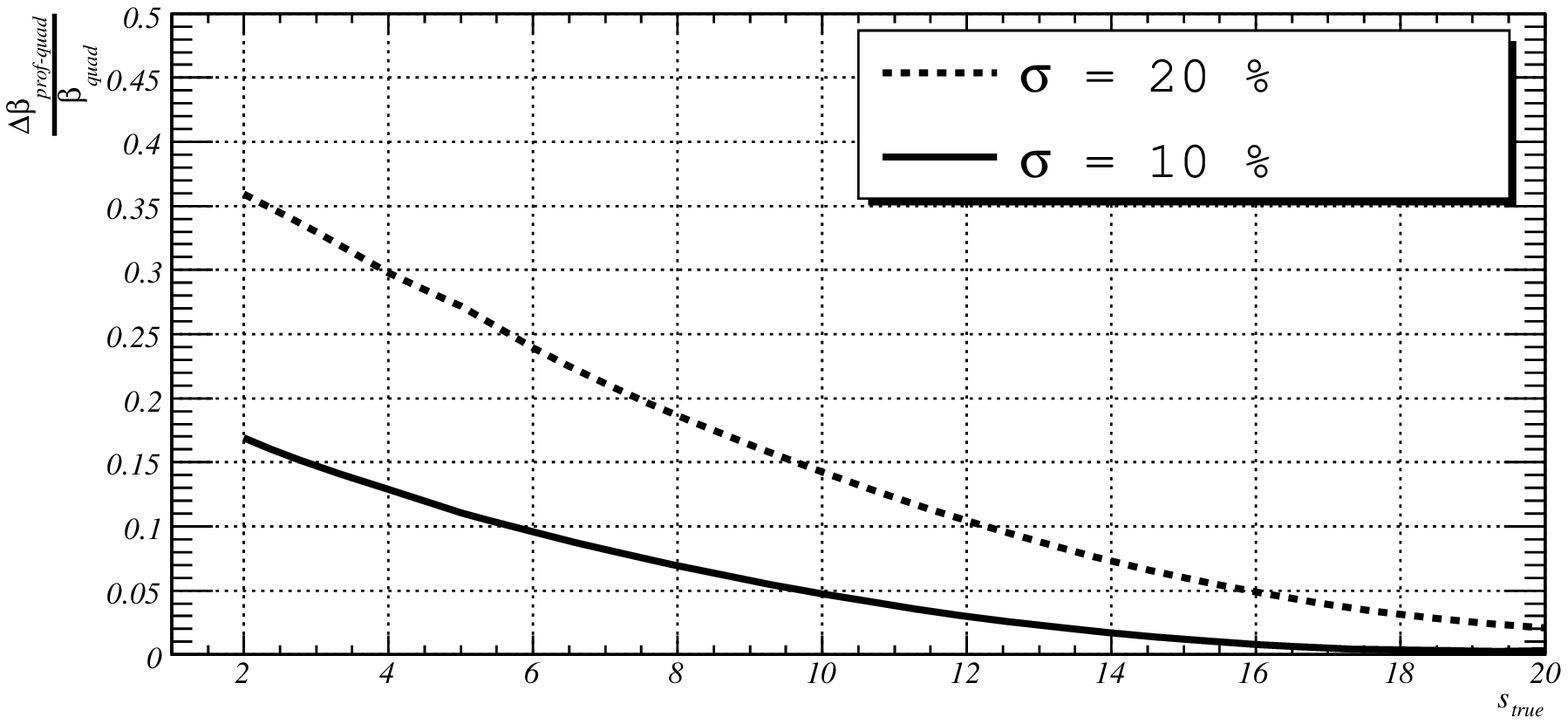,height=10cm,width=12cm} 
\caption{The relative difference between the power of $\chi^2_{prof}$ and $\chi^2_{add}$ as a function of true signal parameter. Here the true background was assumed to be $b_{true} = 10.$}%
\label{fig:pow}
\end{figure}
\end{center}

\noindent
Figures \ref{fig:cov}, \ref{fig:cov2} and \ref{fig:pow} exemplify the results. Figure \ref{fig:cov} shows the value of the modified $\chi^2$s as a function of corresponding coverage. Results are shown for a true background of $b_{true} = 10$ and an uncertainty in the background estimate of 20 \%. For this very simple example,  it can clearly be seen that ignoring uncertainties (in this case in the background estimate) leads to a increased rate of false detections with respect to the one the experimenter intents. The real false detection rate for 99 \% nominal threshold for example is larger by a factor $\sim$3. The effect becomes smaller if one decides to include the additional uncertainties in one of the two ways described in equation \ref{eq:chi11} and \ref{eq:chi21}. Using the latter for example the false detection rate  increases by 50 \% with respect to that nominally required.\\  

\noindent
Though we are assuming a background of $b_{true} = 10$, part of the found difference could be due to the fact that we use $\chi^2$ statistics for a Poisson process. We therefore include the case where we assume a strictly Gaussian measurement process (see fig. \ref{fig:cov}, right panel). The difference between the methods becomes less pronounced, but is still large.\\ 

\noindent
In figure \ref{fig:cov2} we show results for smaller uncertainties in the background estimate (10 \%). As intuitively expected, the impact of the method chosen to calculate the significance becomes less important. If we consider a truely Gaussian process (right panel) for smaller uncertainties both the method using quadratic addition and profiling give results compatible with a nominal $\chi^2$ distribution.

\noindent
The complete measurement process consists of a measurement of background and a measurement of signal events. A complete $\chi^2$ is therefore:
\begin{equation}
\label{eq:chi22}
\chi^2_{emp} = \frac{(n_{obs}- b_{true})^2}{b_{true}} + \frac{(b_{est} - b_{true})^2}{\sigma_{b}^2}
\end{equation}
and $\chi^2_{emp} \sim \chi^2 (2 d.o.f.)$. This quantity is included in the figures.

\noindent
In figure \ref{fig:pow} we show the relative difference in power between the quadratic addition and the profile method. For large signals the power non unexpectedly approaches one, i.e. the method used to calculate the test statistics does not matter. For low signals however one sees that the power of the profile method is up to 35 \% larger than for the method of adding the uncertainty in quadrature.

\subsection{Remark on the ensemble of experiment replica}
In our simple toy model the set of measurements is $(n,b_{est})$ and since we know the distribution of these measurements the ensemble of experiment replica is easily constructed. Under more realistic experimental conditions many different both correlated and uncorrelated nuisance parameter might need to be considered. This might become computationally very cumbersome, for example if full detector simulations need to be employed. Sometimes even uncertainties in theoretical estimates have to be considered. It seems doubtful, though possibly the only feasible way, to treat these as random variables.

\section{Summary \& Conclusions}
Two subjects have been discussed in this note:
\begin{itemize}
\item{\it The interpretation of ``sensitivity''} \\

\noindent
Usually estimates of sensitivity are based on an average experimental result. For a Gaussian distribution of estimates, this implies that if $\Theta_{13} = \Theta_{13}^{sens}$ the probability for claiming discovery will be only 50 \%. This is a well known, but often ignored fact. In our experience, many physicists have the notion that if the value of the true value of the parameter is indeed equal to the sensitivity, then it should be very likely that a discovery will be made.\\ 

\noindent
If the distribution of estimates is not Gaussian, in general no statement about the probability of detection is made if only the average experimental result is presented. Consequently, statements about the probability for detection should be included in presentation of sensitivity estimates. They can be calculated if the distribution of estimates is known or can be simulated.\\

\item{\it Effect of uncertainties}\\

\noindent
The results of the toy model calculation show that if uncertainties in nuisance parameters are included into the calculation of sensitivities (and measurement results) extra care has to be taken to make sure statistical statements (like the significance of a discovery, or confidence level of an interval) are still valid. The largest mistake is not surprisingly made if the uncertainties are ignored. The choice of method to include the systematics furthermore affects the probability of making a discovery. In addition,  in presence of sizable instrumental uncertainties, the ensemble of experiments for calculating significance and power needs to be carefully defined.\\
\end{itemize}

\noindent
When comparing sensitivity estimates for different experiments and experimental configurations, differences therefore certainly could arise from the way the uncertainties are included (if at all) in the calculations. It seems obvious, that sensitivity curves need to be compared at the same ``real'' significance level and at the same ``real'' probability for discovery, whereas they are usually compared for the same nominal significance level and under the assumption that the probability for discovery will be always 50 \%.\\

\noindent
The toy model presented here is a crude simplification of the actual experimental situation where many measurement bins and different types of correlated and uncorrelated uncertainties have to be considered. For example, a generalization of the profiling method to a more realistic experimental situation, including many bins and correlated systematic uncertainties, is given in \cite{Fogli:2002pt}. 
The results presented in figures \ref{fig:cov}, \ref{fig:cov2} and figure \ref{fig:pow} should therefore rather serve as an inspiration for detailed studies of realistic experimental conditions (see for example \cite{Schwetz:2006md}). If possible, the statistical quantities should be studied using Monte Carlo simulations of many replica of the experiment under study. Application of Monte Carlo simulations does not only yield information on the correct false detection rates but also on the probability of detection, which is of obvious importance for assessment of sensitivities of future experiments.

\section{Acknowledgments}
This research was supported by the Swedish Research Council under grant 621-2004-2196. We thank an anonymous referee for useful suggestions.


\begin{thebibliography}{00}
\bibitem{ISS:2007} International scoping study, http://www.hep.ph.ic.ac.uk/iss/

\bibitem{Sinervo:2003wm}
  P.~Sinervo,
{\it In the Proceedings of PHYSTAT2003: Statistical Problems in Particle Physics, Astrophysics, and Cosmology, Menlo Park, California, 8-11 Sep
2003, pp TUAT004}.

\bibitem{Komatsu:2002sz}
  M.~Komatsu, P.~Migliozzi and F.~Terranova,
  J.\ Phys.\ G {\bf 29}, 443 (2003)
  [arXiv:hep-ph/0210043].
\bibitem{Indumathi:2006gr}
  D.~Indumathi, M.~V.~N.~Murthy, G.~Rajasekaran and N.~Sinha,
  Phys.\ Rev.\  D {\bf 74}, 053004 (2006)
  [arXiv:hep-ph/0603264].
\bibitem{Conrad:2006} J.~Conrad, Invited contribution to 8th International Workshop on Neutrino Factories and Superbeams (NuFact 06), Irvine, USA, August 2006, http://nufact06.physics.uci.edu/Workshop/Program/Plenary.aspx
\bibitem{Schwetz:2006md}
  T.~Schwetz,
  Phys.\ Lett.\  B {\bf 648} (2007) 54
  [arXiv:hep-ph/0612223].
\bibitem{Punzi:2003bu} G.~Punzi,
{\it In the Proceedings of PHYSTAT2003: Statistical Problems in Particle Physics, Astrophysics, and Cosmology, Menlo Park, California, 8-11 Sep
2003, pp MODT002}
  [arXiv:physics/0308063].
\bibitem{Conrad:2002kn} J.~Conrad, \emph {et. al.}, Phys.\ Rev.\  D {\bf 67} (2003) 012002 [arXiv:hep-ex/0202013].
\bibitem{Rolke:2004mj}
  W.~A.~Rolke, A.~M.~Lopez and J.~Conrad,  Nucl.\ Instrum.\ Meth.\  A {\bf 551} (2005) 493  [arXiv:physics/0403059].

\bibitem{Burguet-Castell:2005pa} J.~Burguet-Castell, D.~Casper, E.~Couce, J.~J.~Gomez-Cadenas and P.~Hernandez,
  Nucl.\ Phys.\  B {\bf 725} (2005) 306
  [arXiv:hep-ph/0503021].
\bibitem{Huber:2002mx}
  P.~Huber, M.~Lindner and W.~Winter,
  Nucl.\ Phys.\  B {\bf 645} (2002) 3
  [arXiv:hep-ph/0204352].
\bibitem{Barger:2006vy}
  V.~Barger, M.~Dierckxsens, M.~Diwan, P.~Huber, C.~Lewis, D.~Marfatia and B.~Viren,
  Phys.\ Rev.\  D {\bf 74}, 073004 (2006)
  [arXiv:hep-ph/0607177].
\bibitem{Fogli:2002pt}
  G.~L.~Fogli, E.~Lisi, A.~Marrone, D.~Montanino and A.~Palazzo,
  Phys.\ Rev.\  D {\bf 66} (2002) 053010
  [arXiv:hep-ph/0206162].

\end{thebibliography}
\end{document}